\documentclass[review]{elsarticle}

\usepackage{hyperref}

\usepackage{amsmath}
\usepackage{graphicx}
\usepackage{subfigure}
\usepackage{multicol} 
\usepackage{epstopdf}
\usepackage{epsfig}
\usepackage{multirow}
\usepackage{bm}
\usepackage{array}
\usepackage{color}
\usepackage{float}
\usepackage{amsfonts}
\usepackage{caption}
\usepackage{times}

\usepackage{rotating}
\usepackage{bbm}
\usepackage{latexsym}

\usepackage{verbatim}

\begin{document}

\begin{frontmatter}

\title{Reconstruction of Natural Visual Scenes from Neural Spikes with Deep Neural Networks}



\author{Yichen Zhang\fnref{pku}}

\author{Shanshan Jia\fnref{pku}}

\author{Yajing Zheng\fnref{pku}}

\author{Zhaofei Yu\fnref{pku}}

\author{Yonghong Tian\fnref{pku}}

\author{Siwei Ma\fnref{pku}}

\author{Tiejun Huang\fnref{pku}}

\author{Jian K. Liu\fnref{uol}}

\fntext[pku]{National Engineering Laboratory for Video Technology, School of Electronics Engineering and Computer Science, Peking University, Beijing, and Peng Cheng Laboratory, Shenzhen, China}

\fntext[uol]{Centre for Systems Neuroscience,  
        Department of Neuroscience, Psychology and Behaviour, University of Leicester, Leicester, UK}

\begin{abstract}
Neural coding is one of the central questions in systems neuroscience for understanding of how the brain processes stimulus from the environment, moreover, it is also a cornerstone for designing algorithms of brain-machine interface, where decoding incoming stimulus is highly demanded for better performance of physical devices. Traditionally researchers have focused on functional magnetic resonance imaging (fMRI) data as the neural signals of interest for decoding visual scenes. However, our visual perception operates in a fast time scale of millisecond in terms of an event termed neural spike. There are few studies of decoding by using spikes. Here we fulfill this aim by developing a novel decoding framework based on deep neural networks, named spike-image decoder (SID), for reconstructing natural visual scenes, including static images and dynamic videos, from experimentally recorded spikes of a population of retinal ganglion cells. The SID is an end-to-end decoder with one end as neural spikes and the other end as images, which can be trained directly such that visual scenes are reconstructed from spikes in a highly accurate fashion. Our SID also outperforms on the reconstruction of visual stimulus compared to existing fMRI decoding models. In addition, with the aid of a spike encoder, we show that SID can be generalized to arbitrary visual scenes by using the image datasets of MNIST, CIFAR10, and CIFAR100. Furthermore, with a pre-trained SID, one can decode any dynamic videos to achieve real-time encoding and decoding of visual scenes by spikes. Altogether, our results shed new lights on neuromorphic computing for artificial visual systems, such as event-based visual cameras and visual neuroprostheses. 
\end{abstract}


\end{frontmatter}


\section{Introduction}
In everyday life, various types of sensory information are processed by our brain through different sensory modalities, which are then processed to generate a series of behavior reactions. Neuronal networks in the brain, then, play an essential role of efficient and powerful computation where a sequence of input-output mappings is carried out by single and networked neurons. At the level of the system, single neurons receive and respond to input stimuli by changing their membrane potential to generate a sequence of fast events, termed neural spikes.
Thus, spikes have been suggested as a fundamental element to represent input-output neural computation of visual scenes~\citep{chichilnisky2001a, gollisch2008rapid}. Such observations lead to a central question for systems neuroscience on the sensory system: how neurons represent the input-output relationship between stimuli and their spikes~\citep{Rieke1996, Simoncelli2001Natural, wu2006complete}. This question, formulated as neural coding, consists of two essential parts, encoding and decoding. For the visual system, it is to understand how visual scenes are represented by neural spiking activity, and how to decode them to reconstruct the input visual scenes. 

The retina serves as a useful system to study these questions. Visual scenes are projected into the eyes, where the retinal neuronal network processes the input by a few types of neurons, but the only output neurons are the retinal ganglion cells (RGCs). Therefore, all information of visual scenes is encoded by RGCs that produce a sequence of action potentials or spikes, which are transmitted via the optic nerve to the downstream brain regions. Essentially, visual scenes are represented by spikes of a population of RGCs. One expects that encoding and decoding of visual scenes from RGC spikes can be formulated into a closed form where a computational model can be developed to set up a mapping between visual scenes and RGC spikes. Indeed, much effort has been given to study the encoding part of RGCs. Various neuroscience mechanisms have been identified in understanding the retinal computation of visual scenes by its neurons and neural circuitry~\citep{Jadzinsky2013,grimes2018parallel, Brien2018plasticity, rivlin2018flexible, demb2015functional, Gollisch_2010}. In addition, for understanding the encoding principles of the retina, a number of models are developed based on different properties of neurons and neural circuits in the retina~\citep{meyer2017models, Pillow2008Spatio, yan2017revealing, yan2018revealing, Yu2020}. 

Decoding of RGC spikes to obtain the essential information of visual scenes is processed by the downstream neurons from the lateral geniculate nucleus to the visual cortex, which involves feedbacks from the higher part of the cortex for cognition. However, as the retina does not receive these feedbacks, the RGC spikes can be thought as a minimal computational device to represent visual information as a whole. Therefore, it is suitable to develop a decoding model that can reconstruct visual scenes from the RGC spikes directly~\citep{gollisch2008rapid, botella2018nonlinear}.  This is in particular important for neuroengineering, such as neuroprosthesis, where efficient algorithms of decoding neuronal signals for controlling physical devices are highly demanded~\citep{gilja2012high}. In general, an ideal neuronal decoder should be able to read out and reconstruct stimulus from neural responses. The existing methods for visual scenes reconstruction are mainly divided into two types. The first one is to decode the stimulus category: choosing a stimulus according to neural responses, then classifying neural signals to find the corresponding stimulus in a candidate set~\citep{wen2018neural}. The second type is to reconstruct the original stimulus directly by using neural responses, i.e., obtaining every pixel of visual scenes from the neural signal. It is obvious that the second category is more challenging~\citep{nishimoto2011reconstructing}.

Reconstruction of visual scenes has been studied over many years. The neural signals of interest can be functional magnetic resonance imaging (fMRI) activity~\citep{wen2018neural, nishimoto2011reconstructing, thirion2006inverse, naselaris2009bayesian, qiao2018accurate}, neural spikes in the retina~\citep{gollisch2008rapid, botella2018nonlinear, marre2015high, Parthasarathy2017} and lateral geniculate nucleus~\citep{Stanley1999}, neural calcium imaging signals in the visual cortex~\citep{garasto2018visual}.
However, the decoding performance of current methods is rather low for natural scenes, in particular for dynamical videos, which can be seen from some typical examples of the videos reconstructed from fMRI signals~\citep{wen2018neural, nishimoto2011reconstructing}. For the retina, one would expect to decode visual scenes by using the spiking responses of a population of RGCs in a complete way as these spikes are the only output of the eyes. Decoding of visual scenes, at least for static natural images, is possible when sampling the whole image with a number of RGCs~\citep{gollisch2008rapid}. Yet, it remains unclear how to deal with the dynamic natural scenes, where the temporal complexity of stimulus information has a strong coupling with the temporal adaption of neurons~\citep{Liu_2015}.



\begin{figure*}[t]
	\begin{center}
		\includegraphics[width=0.9\linewidth]{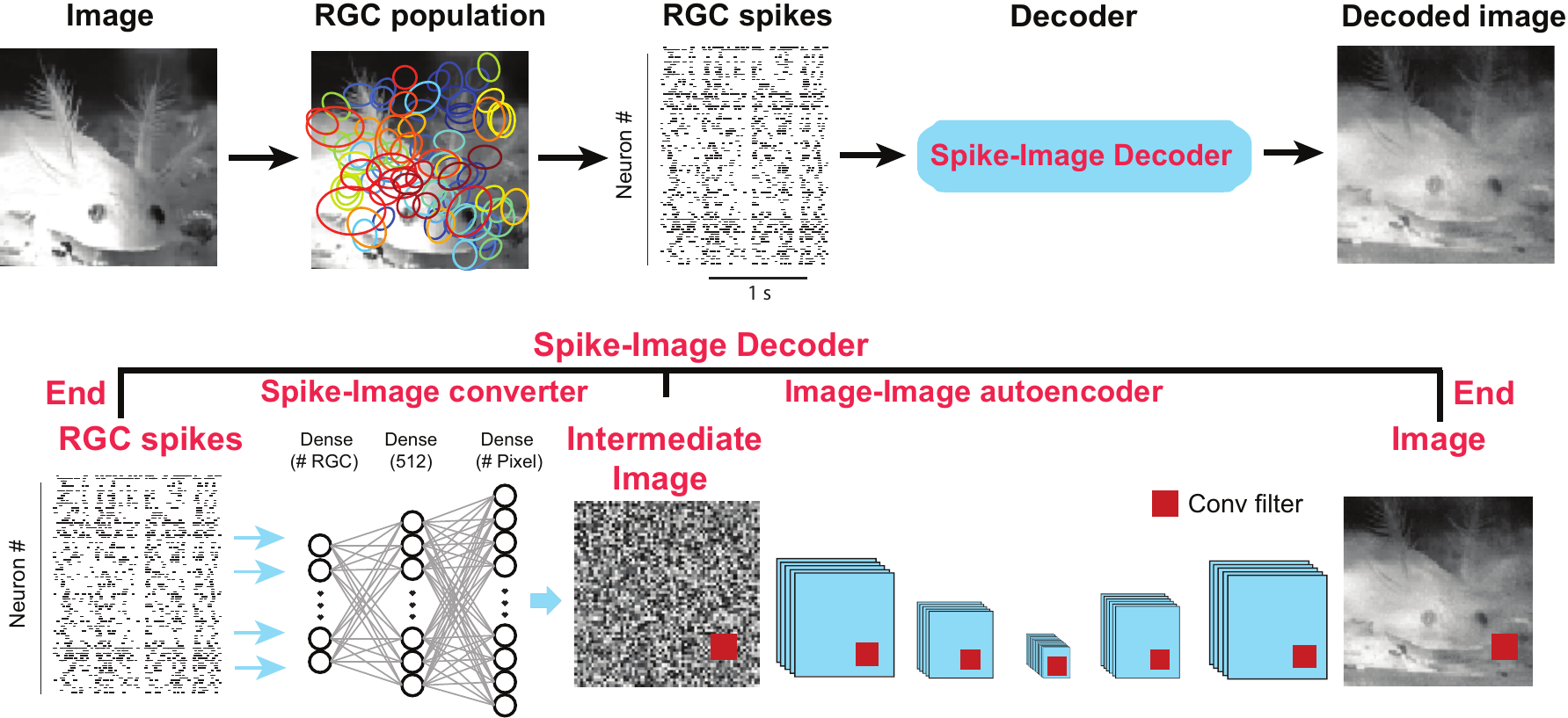}
	\end{center}
	\caption{ 
		Illustration of the spike-image decoder model. (Top) Workflow of decoding visual scenes. Here a salamander swimming video was presented to a salamander retina to stimulate a firing of spike trains in a population of RGCs. The spikes are used to train the decoder for the given video to reconstruct the visual scene. Receptive fields of ganglion cells are mapped onto the image. Each coloured circle is an outline of receptive field. (Bottom) Spike-image decoder is an end-to-end decoder with two stages: spike-image converter used to map a population of spikes to a pixel-level intermediate image, and image-autoencoder for mapping every pixel to the target pixels in the desired images.   
	} 
	\label{fig:decoder}
\end{figure*}

In this study, we propose such a general-purpose decoding framework, termed spike-image decoder (SID), that performs an end-to-end decoding process from neural spikes to visual scenes, based on deep learning neural networks. The proposed SID can achieve state-of-the-art performance, compared to previous studies, for reconstructing natural visual scenes, including both static images and dynamic videos, from spikes of a population of RGCs recorded simultaneously in the isolated animal retina. The workflow of the SID is illustrated in Figure~\ref{fig:decoder}. When a large population of RGCs is recorded simultaneously, and their spikes are extracted, a spike-image converter based on a neural network is used to map the spikes of every RGC to intermediate images at the pixel level. After that, an autoencoder-type deep learning neural network is applied to spike-based intermediate images to match original stimulus images for every pixel. 

Essentially, the SID has two stages with one as spike-image converter and another one as image-image autoencoder. Most of the previous studies focused on the first stage, such that the spikes are directly mapped to obtain the image pixels, where, traditionally, a decoder can be optimized by some statistical models in a linear or nonlinear fashion~\citep{gollisch2008rapid,botella2018nonlinear,nishimoto2011reconstructing,thirion2006inverse,naselaris2009bayesian,marre2015high,Stanley1999}. A recent study used a separate autoencoder network as the second stage to enhance the quality of images~\citep{Parthasarathy2017}, which essentially employed a similar network model dedicated for image denoising and enhancement~\citep{yue2018simple}. Here, we systematically tested different training approaches of neural networks from spikes to images, and found the SID can achieve similar performance for static images across three types of loss functions, which is consistent with those previous studies, however for dynamic videos, the direct training from spikes to images shows an outstanding performance. 

We tested the performance of SID on experimentally recorded RGCs spikes, and found it can obtain state-of-the-art performance for reconstruction of natural visual scenes of static image and dynamic video. Our SID can also be used to decode the fMRI signal, and has a better performance on the reconstruction of visual stimulus compared to the existing state-of-the-art fMRI decoding models. Furthermore, we found that the computational ability of SID can be generalized to arbitrary visual scenes with the aid of an encoding model that can simulate spikes for any given images. By using the image datasets of MNIST, CIFAR10, and CIFAR100, the SID can reconstruct natural images with high precision. In addition, with a pre-trained SID model, one can reconstruct any given video scenes to realize real-time encoding and decoding of dynamic visual scenes as an artificial vision system.

\section{Methods}

\subsection{Spike-image decoder}

Traditionally, it has been shown that deep neural networks can extract low-level features in earlier layers, and more semantic features in later layers, which is similar to the information processing of the visual cortex in the brain~\citep{zeiler2014visualizing, Yamins2016Using}. Here we used a similar deep neural network to reconstruct natural scenes from the retinal spikes.
Our decoding model consists of two parts: 1) spike-to-image converter and 2) image-to-image autoencoder. The spike-to-image converter is able to map every spike to every pixel to get intermediate images, which plays a role of nonlinear re-sampling, and the image-to-image autoencoder is used to match all pixels to those of the target images of visual scenes.  

\subsubsection{Spike-to-image converter}
The spike-to-image converter is a three-layer fully-connected neural network, similar to a multilayer perceptron (MLP). The first layer receives the spikes of all RGCs as input such that the number of neurons of the first layer is matched to the number of RGCs used. The second layer is a hidden layer, which consists of 512 neurons. ReLU is used as an activation function of the hidden layer. The output layer has the same number of neurons as the pixel number of the input image, in our case, 4096 neurons were used since all stimulus images are in 64*64 pixels. ReLU as activation function is used for the third layer, and we can get all pixel values of intermediate image from this output layer. 

\subsubsection{Image-to-image autoencoder}
The image-to-image autoencoder is a typical deep autoencoder based on convolutional neural network (CNN), where the whole information processing procedure can be split into two phases. In the first phase, convolution and down-sampling are used to process and decrease the size of the input image. This phase contains four convolutional layers, the kernel sizes of these four layers are (64,7,7), (128,5,5), (256,3,3), (256,3,3) and the stride sizes are (2,2) for all these four layers. After this phase, the most important components of the input image are kept, and noise and redundant contents are filtered. In the second phase, convolution and up-sampling are used to process the image, which recovers the texture of the down-sampled image while increasing the size of the down-sampled image. The upsampling phase also consists of four convolutional layers, the kernel sizes are (256,3,3), (128,3,3), (64,5,5), (3,7,7) and the stride sizes are (1,1) for all these layers. Up-sampling layers with size (2,2) are added before these four convolutional layers. ReLU is used as an activation function in the whole Image-to-image autoencoder.

\subsubsection{SID model architecture}

Given an input image $\textbf{I}$, it will trigger a response $\textbf{s}=\{s_1, s_2 \dots s_n\}$ on the retinal ganglion cells, here we use rate coding such that $s_i$ represents the spike count of each RGC within a time bin depending on the sampling rate of visual scenes. As in experimental recordings, a typical 30 Hz refreshing rate is used, so the time bin is about 33 ms. Then the triggered responses are first fed into the spike-to-image MLP converter, which outputs an intermediate image $\textbf{O}_1=f_1(\textbf{s})$ from RGC responses. Then the image-to-image CNN autoencoder takes the intermediate image as input, and is able to map it to match the target image, so one can get a clear and refining reconstruction result $\textbf{O}_2=f_2(\textbf{O}_1)$. These two parts are all implemented with deep neural networks, so end-to-end training can be used to train both networks. 

Three different ways of defining loss function were explored in this study. In the first one, one can use a multi-task loss function to make sure both stages have the capability to decode natural scenes from spikes, so that we can get a high quality reconstruction result from our decoding model. Here, $\lambda_1$ and $\lambda_2$ are weights of these two parts mean square error loss, $\textbf{L}_1: Loss = \lambda_1 || \textbf{O}_1 - \textbf{I} || + \lambda_2  ||\textbf{O}_2 - \textbf{I}||$. 
In the second one, we optimize each stage of decoding separately such that
$\textbf{L}_2: (Loss_1 =\lambda_1 || \textbf{O}_1 - \textbf{I} ||, \, \, \,\,  Loss_2 = \lambda_2  ||\textbf{O}_2 - \textbf{I}||).
$
The third one uses one loss for the whole process of two stages as following:
$
\textbf{L}_3: Loss=\lambda || \textbf{O} - \textbf{I} ||
$, where there is only one image output $\textbf{O}$ as the final reconstruction to compare the target image such that one error signal stays in both stages. In this way, the intermediate image is like noise without visible structures as in Figure~\ref{fig:decoder}. Three loss functions were compared in details.

The SID was implemented with Keras deep learning library, and Tensorflow as backend was used. During training for both spike-to-image converter and image-to-image autoencoder, batch normalization using 256 as batch size for both converter MLP and autoencoder CNN was used to achieve better performance~\citep{ioffe2015batch} and dropout~\citep{hinton2012improving} was used to deal with overfitting. Adam method was used to train the model. The code is available online at https://sites.google.com/site/jiankliu.

\subsection{Experimental spike data}

Our decoder was used to reconstruct natural visual scenes, including both static image and dynamic video, from spikes of a population of RGCs recorded simultaneously in isolated retinas of salamanders. Datasets of spikes recorded and stimulus images and movies used in this study can be found publicly~\citep{Onken_2016}. Experimental details can be found in previous studies~\citep{Onken_2016, Liu2017}, Briefly, 
neural data had been collected in isolated retinas obtained from axolotl salamanders. The retina was placed onto a 60- or 252-channel multielectrode array in a recording chamber. Visual stimuli were displayed on an OLED monitor that was projected onto the photoreceptor layer through a telecentric lens above the retina.

\subsubsection{Image stimulus}
The static images were taken from a dataset used previously~\cite{Onken_2016, Liu2017}, which is part of the McGill Calibrated Colour Image Database~\citep{Olmos2004}, including 300 natural images with a size of 64*64 pixels. Briefly, each image covers 1920*1920 $\mu$m area on the retina, in which there are 80 RGCs recorded. Each individual image was presented for 200 ms and then followed by an 800 ms empty display in a pseudo-random sequence. 
For each image to each RGC, the number of spikes during 300 ms in one trial was collected. In addition, we averaged the spike counts among 13 trials for each image. Finally, we got the spike counts of a population of 80 RGCs for one image as the input data for reconstruction. Among all 300 images, 270 images were used to train the decoder model and the remaining 30 images were used as the test set. 

\subsubsection{Video stimulus}
The video data consists of salamander swimming which includes 1800 frames as the same dataset in~\citep{Onken_2016}. Each frame covered a total area of 2700*2700 $\mu$m on the retina with a spatial resolution of 360*360 pixels, which was down-sampled into 90*90 pixels, and cropped into 64*64 pixels around the central region of images, in which 90 RGCs responses were collected. A segment of the video was presented at a frame rate of 30 Hz. For the recorded responses of each RGC, one can bin them into 33 ms that is same with the presentation time of each frame. As a result, a spike train was collected with 1800 counts for each RGC in response to the video. Eventually for the population of RGCs, a 90*1800 spike count matrix, averaged over 31 trials of repeated presentation, were used to reconstruct the video. To train the decoder model, we randomly selected 1620 frames from the video as a training set and the remaining frames as a test set. In this way, the strong temporal correlation within the video can be washed out. 

\subsection{Simulated spike data}

To test our SID model for any arbitrary visual scenes, we used a simple encoder based on the typical linear-nonlinear model~\citep{chichilnisky2001a, Liu2017} to simulate neural responses of a population of RGCs. For comparison, the encoding model is based on the same experimental retinal data used for the analysis of the video. The linear filters are based on the receptive fields of a population of 90 RGCs obtained with white noise analysis~\citep{chichilnisky2001a, Onken_2016, Liu2017} fitted with a 2D Gaussian for each cell. With the given natural scene stimulus (a static image or a frame of dynamic video), each cell implements a linear computation on pixels within its receptive field, and generates a spike count for this stimulus after a rectification nonlinearity.
All these simulated spike counts (modeled responses of 90 RGCs) were then fed into our SID model to reconstruct natural scene stimulus. 

With this retinal encoding model, one can simulate RGC responses under any given stimuli. In this study, simulations were done on three popular image datasets, MNIST, CIFAR10, and CIFAR100~\citep{lecun1998proceedings,krizhevsky2012learning}. MNIST is a dataset of handwritten digits that consists of a training set of 60,000 examples and a test set of 10,000 examples. CIFAR10 and CIFAR100 are datasets used for image classification, consisting of 60,000 color images in 10 and 100 classes respectively. There are 50,000 training images and 10,000 test images in both of CIFAR10 and CIFAR100. All images were first resized to the size of 64*64 pixels, which is comparable to our retinal experimental data. Three channels of each color image were first encoded into spikes by the encoding model separately. The encoded (90,3) spikes are fed into spike-to-image converter to get a color intermediate image, here the weights are shared for all three channels. Then image-to-image autoencoder takes the intermediate image as input to get final reconstructed color image. 

Similarly, for comparison, dynamic videos were also used for testing the SID. With the same encoding model, we used the same movie chips from the fMRI decoding experiments~\citep{nishimoto2011reconstructing}, and obtained a set of RGC spiking responses similar to the biological data. Then the SID model pre-trained with CAFAR10 dataset was used to decode these dynamic videos directly. 

Finally, we implemented real-time coding of arbitrary visual scenes by embedding our models into the power-efficient computing device, here NVIDIA Jetson TX2, which is built around a 256-core NVIDIA Pascal GPU. In order to better take advantage of the GPU to accelerate the computation, we combined the image-spike encoding process into the decoding neural networks. In the real-time coding system, we can choose the input files such as image or video from the embedded device, or directly read the inputs, i.e. real-time videos, from a camera. However, as our SID was pre-trained with $64 \times 64$ pixels as input stimulus and target output, the resolution is not very high such that it misses the fine detailed structure of visual scenes. Therefore, based on the pre-trained SID, we cropped the inputs stimulus into several patches with $64 \times 64$ pixels, and then joined the decoding results after sending these data as a batch into the model simultaneously. Considering the GPU processing speed as well as the resolution, we cropped the input frames as $5\times5$ patches, such that the resolution of input images and decoding results can be increased to $320 \times 320$ in a real-time fashion. Higher resolution with real-time speed can be further achieved depending on the hardware used for this artificial vision system.  
The reconstruction results of the same videos used by the previous study~\citep{nishimoto2011reconstructing} and our real-time video decoding are available online  (https://sites.google.com/site/jiankliu). 

\subsection{Experimental fMRI data}
For comparison, we consider experimental fMRI data recorded in human. The stimuli contain a set of 100 gray-scale handwritten digit images (equal number of 6’s and 9’s) extracted from the MNIST dataset, and the corresponding fMRI recordings were from V1-V3 areas of a single participant as detailed in previous studies~\citep{van2010neural,van2010efficient}. 
We evaluated this fMRI dataset by our SID directly with the same network structure. First, we convolved the stimulus images, here, 6's and 9's, with the same set of receptive field of RGCs used previously to simulated spikes, which is a population of 90 simulated cells, similar to the simulation of MNIST and CIFAR image datasets.

For fMRI data, each stimulus image triggered a response represented by a 3092-dimension vector as activity patterns. For a detailed comparison, we conducted two-fold cross-validation: decoding two sets of neural signals, fMRI and spikes, with two different models. We choose a recent state-of-the-art method termed deep generative multi-view model (DGMM)~\citep{Du2019}, which is developed in the context of fMRI decoding. Thus, both models, DGMM and SID, are used for decoding with both fMRI data and our simulated spikes to see if there is an advantage of neural data type, as well as an advancement of decoding method. Within such a validation, one can observe decoding results in four conditions: decoding fMRI by DGMM (DGMM-fMRI), decoding spikes by DGMM (DGMM-Spike), decoding fMRI by SID (SID-fMRI), and decoding spike by SID (SID-Spike). In all cases, we used 90 images for training and 10 images for test. 

\section{Results}

\begin{figure*}[t]
	\begin{center}
		\includegraphics[width=0.95\columnwidth]{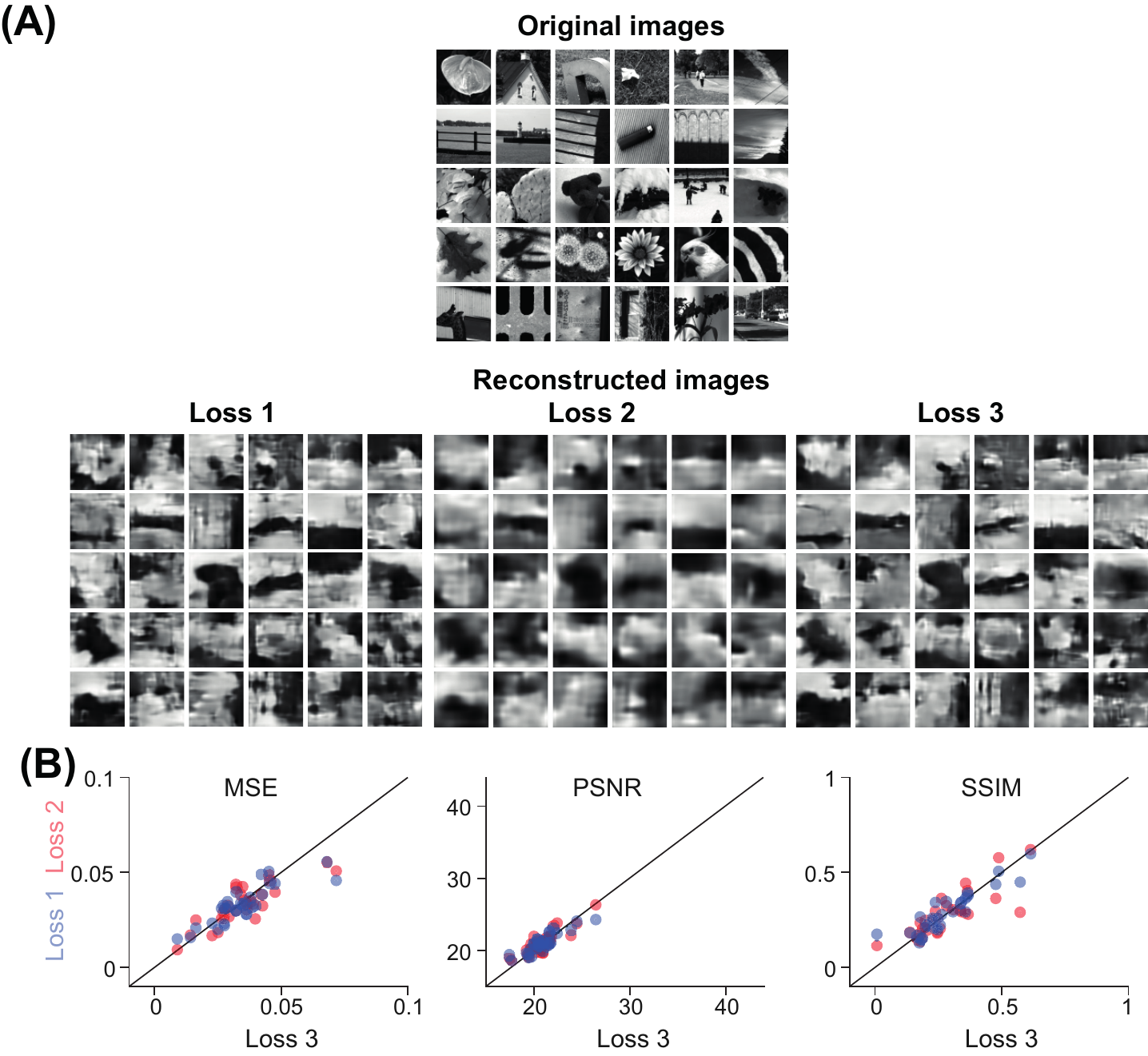}
	\end{center}
	\caption{Reconstructed natural images from experimental RGC spikes with different loss functions. (A) Example original stimulus images in the test data (top) and final reconstructed images (bottom) from RGC spikes with three loss functions. (B) Reconstruction errors measured by MSE, PSNR, and SSIM are compared between Loss 3 and Loss 1 (blue) and Loss 2 (red). 
	}
	\label{fig:img_test}
\end{figure*}

To test the capability of our SID model for reconstruction of visual scenes, we first use our decoding model to reconstruct natural stimulus from biological experimental data that consists of a population of RGCs recorded in salamander retina triggered by natural image and video. To further test the generalization capability of our model, a set of simulated experiments with the encoding model were conducted, in which we reconstructed stimulus images of MNIST, CIFAR10, and CIFAR100 from simulated spikes. In addition, dynamic videos were also tested with simulated RGC spikes and with the SID model trained from CAFAR10. The results show that our SID can decode natural scenes from both biological and simulated spikes with very good precision. 

\subsection{Decoding visual scenes from experimental data}

Here the experimental data are a population of RGCs recorded from salamander retina (see Methods), where the stimuli include static natural images and a dynamic video, and the spike counts of the recorded RGCs are used as the input of our decoding model to reconstruct natural scenes.

\subsubsection{Decoding natural image stimulus}

For natural images, there are 300 images as stimulus and 80 RGCs showing spike responses. For the decoder, the spike counts of 80 RGCs in response to each image are collected at one end, the other end is a training set of 270 images of 64*64 pixels each. For each image, spike-image converter up-sample a population of 80 spike counts into an image with 64*64 pixels to get an intermediate image, which then is mapped to the target training image with the image-image autoencoder.

Figure~\ref{fig:img_test} (A) shows the decoding results on some example test images, where the global contents of images are reconstructed well while some details are missed. 
Depending on the loss functions used for training, intermediate images can be very different (Supplemental Figure 1), however, final reconstructions are similar to each other with some fine differences in the detailed textures. When there are two losses for each stage of the decoder as in $\rm \textbf{L}_1$ and $\rm \textbf{L}_2$, the intermediate image (denoted as $\rm \textbf{O}_1$) is an preliminary image close to the target image. When there is one loss for both stages of the model as in $\rm \textbf{L}_3$, the intermediate image is like noise without any visual structures. 

\begin{figure*}[t]
	\begin{center}
		\includegraphics[width=0.95\columnwidth]{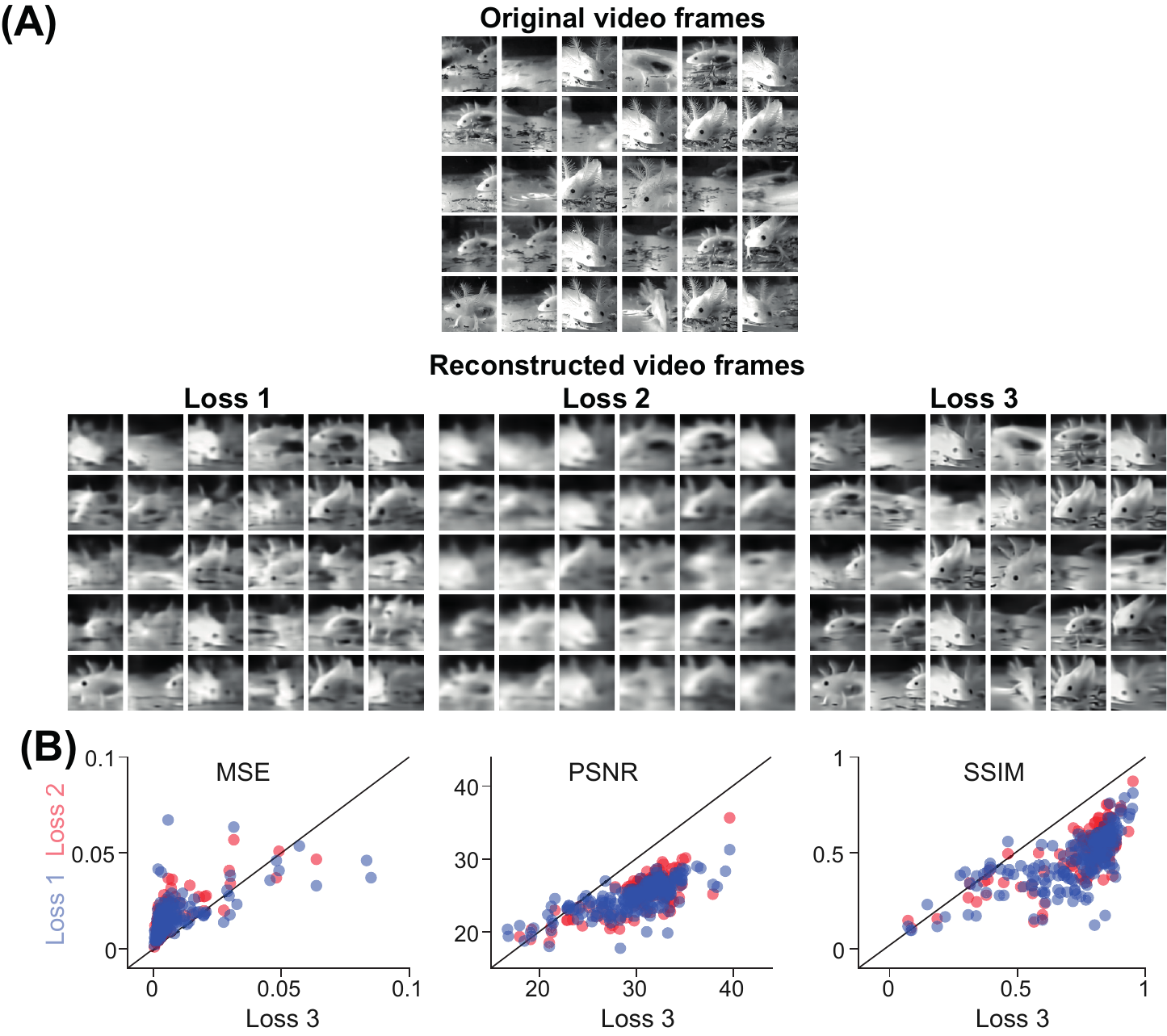}
	\end{center}
	\caption{ 
	Reconstructed video frames from experimental RGC spikes with different loss functions. Similar to Figure~\ref{fig:img_test}, except that here are 30 random test video frames of a continuous video in (A), and a set of 50 random test frame images in (B), where each point is one of the test images.
	}
	\label{fig:move_test}
\end{figure*}

Figure~\ref{fig:img_test} (B) shows the performance of our model characterized by three typical measures of reconstructed images: the mean square error (MSE) that describes the absolute difference of every pixel, the peak-signal-to-noise-ratio (PSNR) that characterizes the global quality, and the structural similarity index measure (SSIM) that captures the details or image distortion, for evaluating the reconstruction results. Whereas there is a variation between different test images, in general, all three loss functions give similar performances in terms of three measures.
However, these measures are not perfect characteristics. For example, the reconstructed images with $\textbf{L}_2$ look as not good as the other two, however, it gives the similar measure errors. Therefore, these measures are a general guideline rather than the precise characteristic, as it seems that there is no perfect measure~\citep{Hore2010,Wanga}.

\subsubsection{Decoding dynamic video stimulus}

So far, most of the decoding studies focus on static natural images, or some simple artificial dynamic scenes~\citep{gollisch2008rapid, botella2018nonlinear}. To study dynamical visual scenes, such as videos with highly complex dynamics in both spatial and temporal domains, one has to, traditionally, understand how to deal with strong temporal adaptation observed in single neurons and neural circuits in the retina~\citep{Liu_2015}. Here with the video stimulus containing 1800 frames, and a population of 90 RGCs spike trains, our SID can overcome this difficulty by training of randomly selected frames out of the whole video, and testing on a subset untrained frames (see Methods).

From the reconstructed sample frames of test dataset shown in Figure~\ref{fig:move_test} (A), one can see that our decoding model obtains high precise results, which can not only reconstruct the global content of each frame, but also some details of images that are missed in previous decoding studies~\citep{botella2018nonlinear,nishimoto2011reconstructing,wen2018neural}. 
The quantified performance is shown in Figure~\ref{fig:move_test} (B).
In contrast to the image results, all three measures, MSE, PSNR, and SSIM, indicate that the best results are obtained with $\textbf{L}_3$. 
Such a good performance can also be seen from the sample images where both global contents and fine details are obtained.

\subsection{Decoding visual scenes from simulated spikes}

\begin{figure}[t]
	\begin{center}
		\includegraphics[width=1\columnwidth]{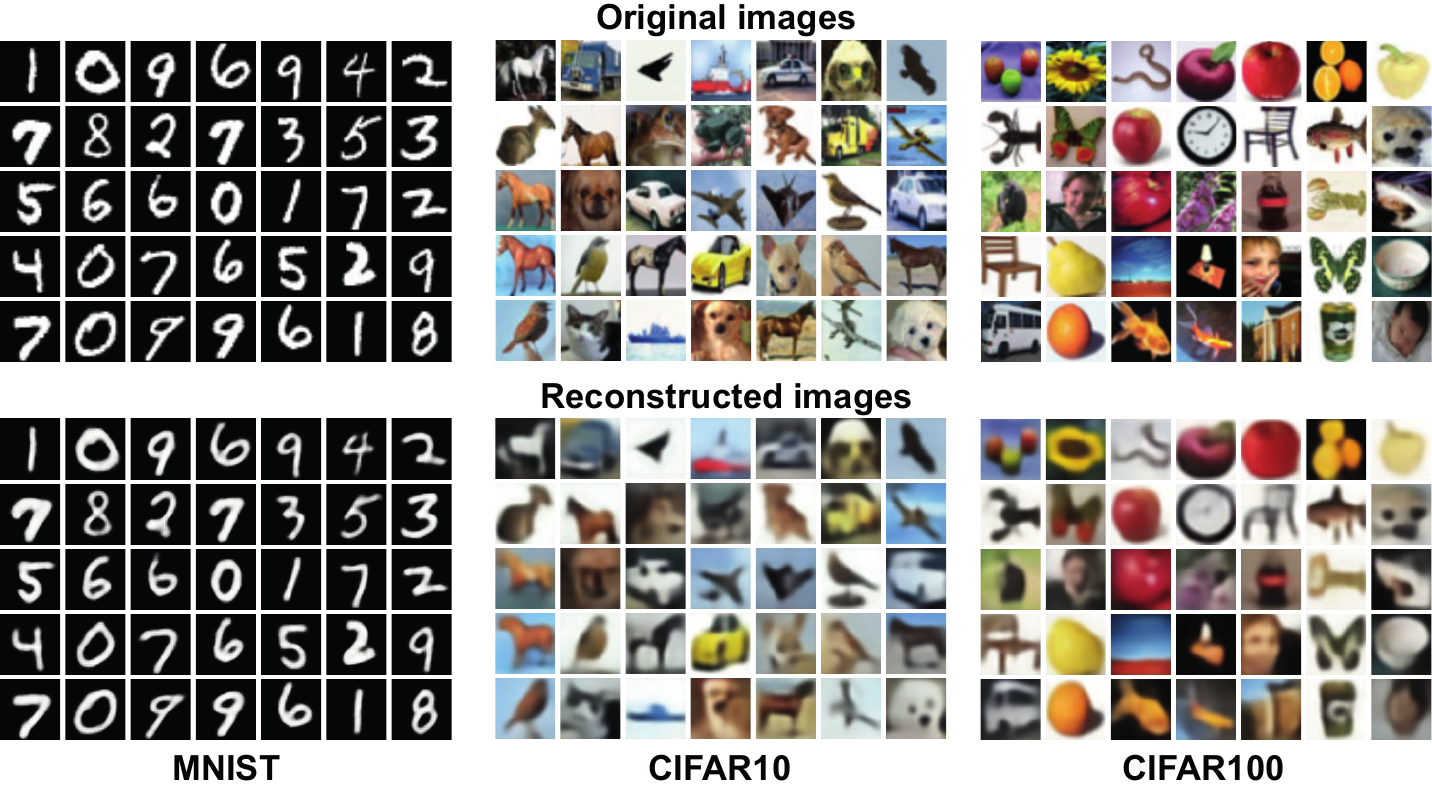}
	\end{center}
	\caption{Reconstructed images from simulated RGC spikes in datasets of NMIST, CIFAR10, and CIFAR100. (Top) Example original images in test dataset. (Bottom) Reconstructed images from the spikes of 90 simulated RGCs. }
	\label{fig:mnist}
\end{figure}

\begin{figure}[t]
	\begin{center}
		\includegraphics[width=0.9\columnwidth]{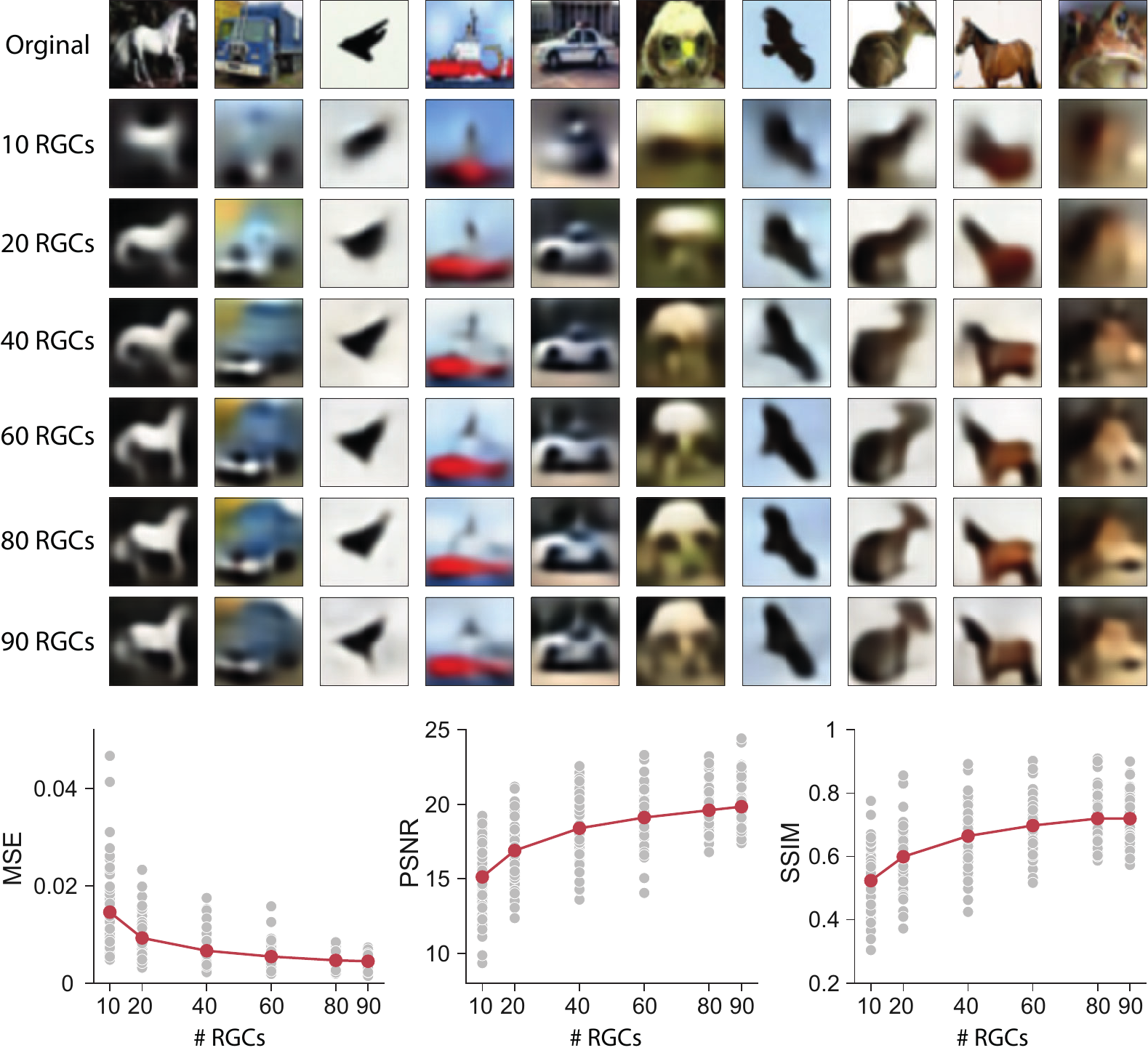}
	\end{center}
	\caption{ Reconstruction with a reduced number of RGC inputs. (Top) Reconstructed example images from CIFAR10 with different subsets of RGCs, from 10 to 90 RGCs. (Bottom) The performance of SID increased with more RGCs. Each gray point represents one of 50 randomly selected test images. Red points are the averaged measures over all test images. 
	}
	\label{fig:less_rgc}
\end{figure}

\begin{figure}[t]
	\begin{center}
		\includegraphics[width=0.95\columnwidth]{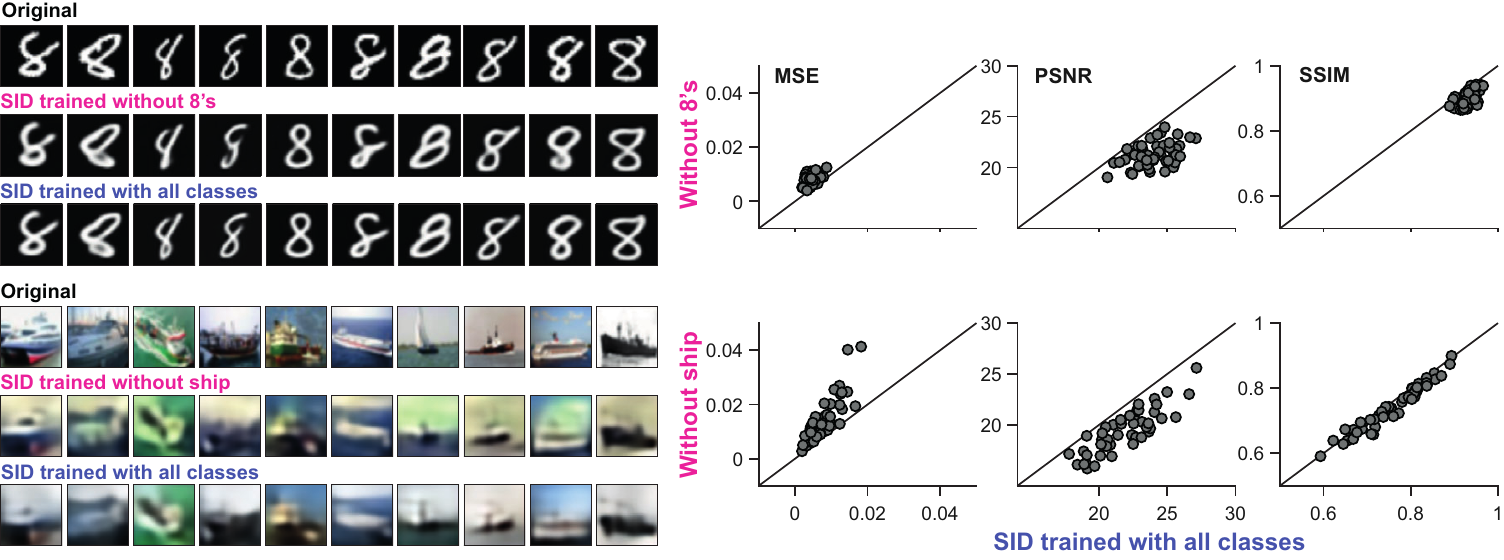}
	\end{center}
	\caption{ SID generalization to the untrained images of one category of MNIST and CIFAR. (Left, top) Original test 8's images from MNIST data, reconstructed example 8's images by the SID trained on other images but without the 8's category, and reconstructed example 8's images by the SID trained on images with all the categories. (Left, bottom) Similar results from CIAFR data, where the ship category is not used for the SID training. (Right) Comparison of generalization ability between the SID  with and without one category of MNIST (top) and CIAFR (bottom). Each gray point represents one of 50 randomly selected test images.} 
	\label{fig:class}
\end{figure}

To further test the generalization capability of our decoding model, we did numerical experiments on simulated RGC data by creating a retinal encoding model based on the typical linear-nonlinear model and biological experimental data to simulate RGC responses (see Methods). In this way, one can simulate RGC spiking responses for any given stimuli. For comparison with experimental RGC data, we simulated the same number of 90 RGCs to obtain spikes.

We first did experiments on three popular image dataset, MNIST, CIFAR10, and CIFAR100. A population of neural spikes was obtained for each image with the encoding model. Then the SID was trained for the training set, and tested on a separate test data. The reconstructed example images are shown in Figure~\ref{fig:mnist} with good quality. In particular, the decoding performance for MNIST is almost perfect.

With the good reconstruction from simulated spikes with a population of only 90 RGCs, one can ask the question that how the decoding performance can be changed by a reduced number of RGCs. Thus, we re-trained the SID with a randomly selected subset of 90 RGCs as the input for the model, as shown in Figure~\ref{fig:less_rgc}. Surprisingly, with only 20 RGC input, the SID can still recover a large part of the information from the original stimulus images. Indeed, the performance measured by MSE, PSNR, and SSIM is systematically decreasing when fewer RGCs are used for input. However, the performance tends to be convergent with 90 RGCs, which is the maximum number of RGCs recorded in our experimental data. It might be possible that the performance can be a little higher when more than 90 RGCs are used. We leave this possibility for the future study when more RGCs can be recorded (See Discussion).

As our decoder reconstructs the image per pixel directly, we expect that the training of SID can be done without any information about the category. Both MNIST and CIFAR datasets are originally used for image classification with labels of category. Thus, we re-trained our SID on these datasets but removing one of the known categories. Then we tested the reconstruction of the missing images from this untrained category. As shown in Figure~\ref{fig:class}, the 8's digital images are reconstructed from the SID that is trained without 8's of digital images. In both training conditions, the sample size of the training set is the same, so that the only difference between the two is the missing category. We found that the 8's can be decoded with a high accuracy, which means that the training of SID does not use any information about the category of the dataset. Similarly, the various ship images from the missing category during the training can also be reconstructed well on the more complex CIFAR dataset as in Figure~\ref{fig:class}. However, the performance characterized by three measures of MSE, PSNR, and SSIM shows that there is a subtle difference between two versions of the test results: the SID trained with all classes has better performance, compared to the SID training without one category. Taken together, these results suggest that our SID has a good generalization ability, and the rich image content of the dataset can improve the decoding ability of the SID.

\begin{figure}[t]
	\begin{center}
		\includegraphics[width=1\columnwidth]{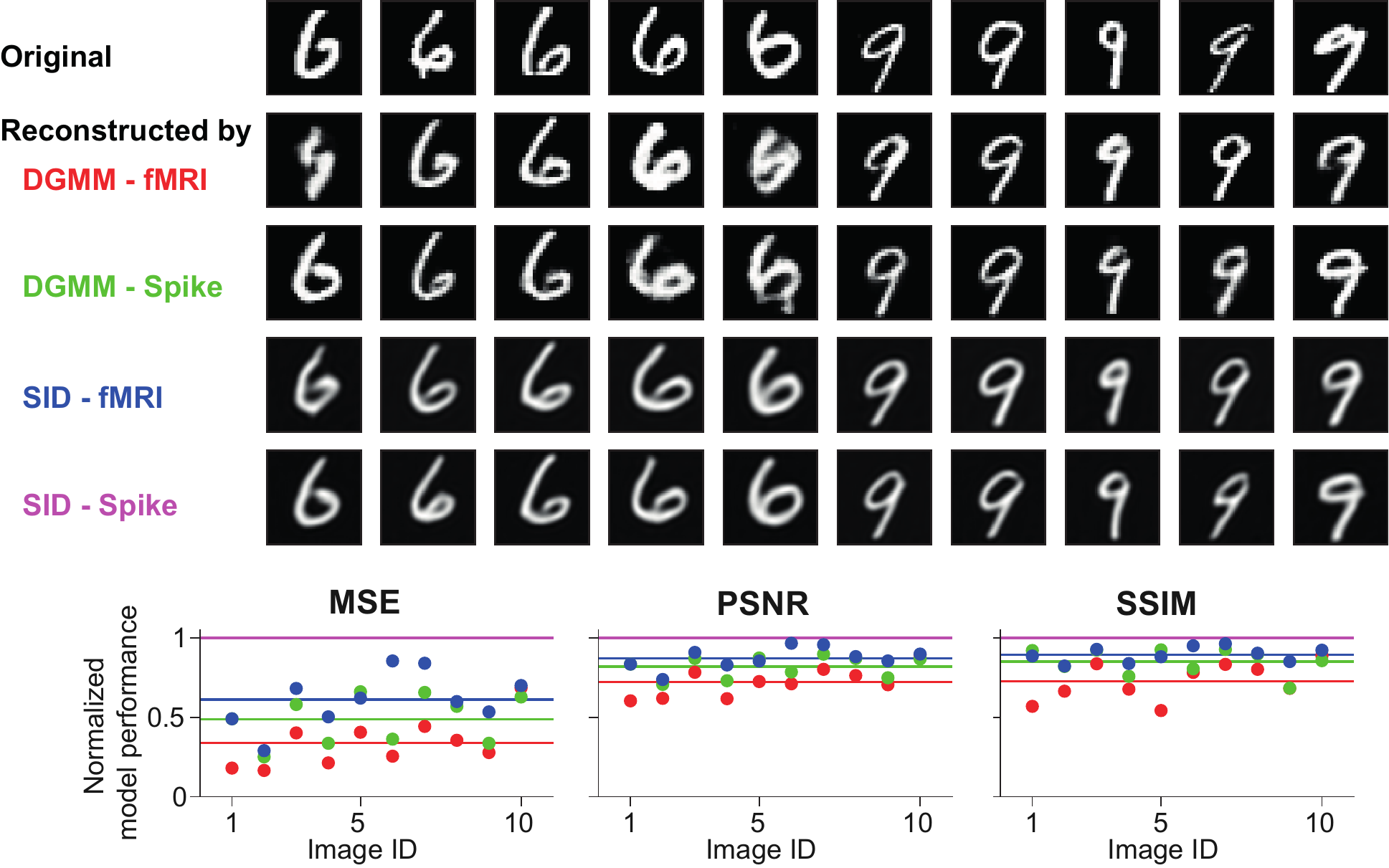}
	\end{center}
	\caption{ Reconstructed stimulus images from fMRI responses and simulated spikes. (Top) All of the 10 original images in the test dataset used for fMRI recordings, and the reconstructed images from fMRI data and the simulated spikes with the decoder of DGMM and SID cross-validated by a pair of model-data test, \textit{e.g.} DGMM-fMRI stands for fMRI decoding by the DGMM model. For comparison, cross-validation is done by using DGMM for simulated spikes and our SID for fMRI data. (Bottom) Normalized model performance measured by MSE, PSNR, and SSIM. The performance of each pair of model-data is normalized by the SID-Spike. Each colored data point corresponds to each test image in the same order of image ID as in the original stimulus images on the top row. Colored solid lines represent the mean values of each error measure: DGMM-fMRI (red), DGMM-Spike (green), SID-fMRI (blue), and SID-Spike (purple). 
	}
	\label{fig:fmri}
\end{figure}

\subsection{Cross-validation with existing methods}

As there is no direct experimental data of neuronal spikes recorded for MNIST so far, we compared our results with the decoding results based on the models developed for fMRI signals. Previous studies evaluated the performance of quite a few decoders based on one particular dataset where a small part of MNIST images, 6's and 9's, were used as stimulus and the fMRI data was collected for human subject~\citep{van2010neural,van2010efficient, fujiwara2013modular, wang2015deep, miyawaki2008visual, wen2018neural, Du2019, qiao2018accurate}. 
We compared our SID with a recent state-of-the-art method, termed Deep Generative Multi-view Model (DGMM), developed for decoding of fMRI data~\citep{Du2019}. For detailed comparison, we used the same fMRI dataset but evaluated the decoding performance based on both fMRI responses and our simulated spikes, as well as two decoding models: DGMM and our SID, such that there are four pairs of model-data test: DGMM-fMRI (\textit{e.g.} fMRI decoding by DGMM), DGMM-Spike, SID-fMRI, and SID-Spike. 

The reconstructed example images are shown in Figure~\ref{fig:fmri}, which can be characterized by MSE, PSNR and SSIM from a single run and a set of 10 runs with different initial conditions of the decoding models (Supplemental Figure 2).
Interestingly, the SID gives a better performance for decoding of fMRI as well.  Compared to DGMM, our SID outperforms on decoding with both fMRI signals and simulated spikes. Figure~\ref{fig:fmri} shows the normalized model performance for each individual test images, where all model-data pairs are normalized by the SID-Spike. In this way, SID-Spike gives a ratio as 1, and other model-data pairs have a performance smaller than 1. In particular, the performance increases in ascending order: DGMM-fMRI $<$ DGMM-Spike $<$ SID-fMRI $<$ SID-Spike. These results suggest that our SID, as a general decoding method, shows a good capability for fMRI decoding as well.


\begin{figure}[t]
	\begin{center}
		\includegraphics[width=0.9\columnwidth]{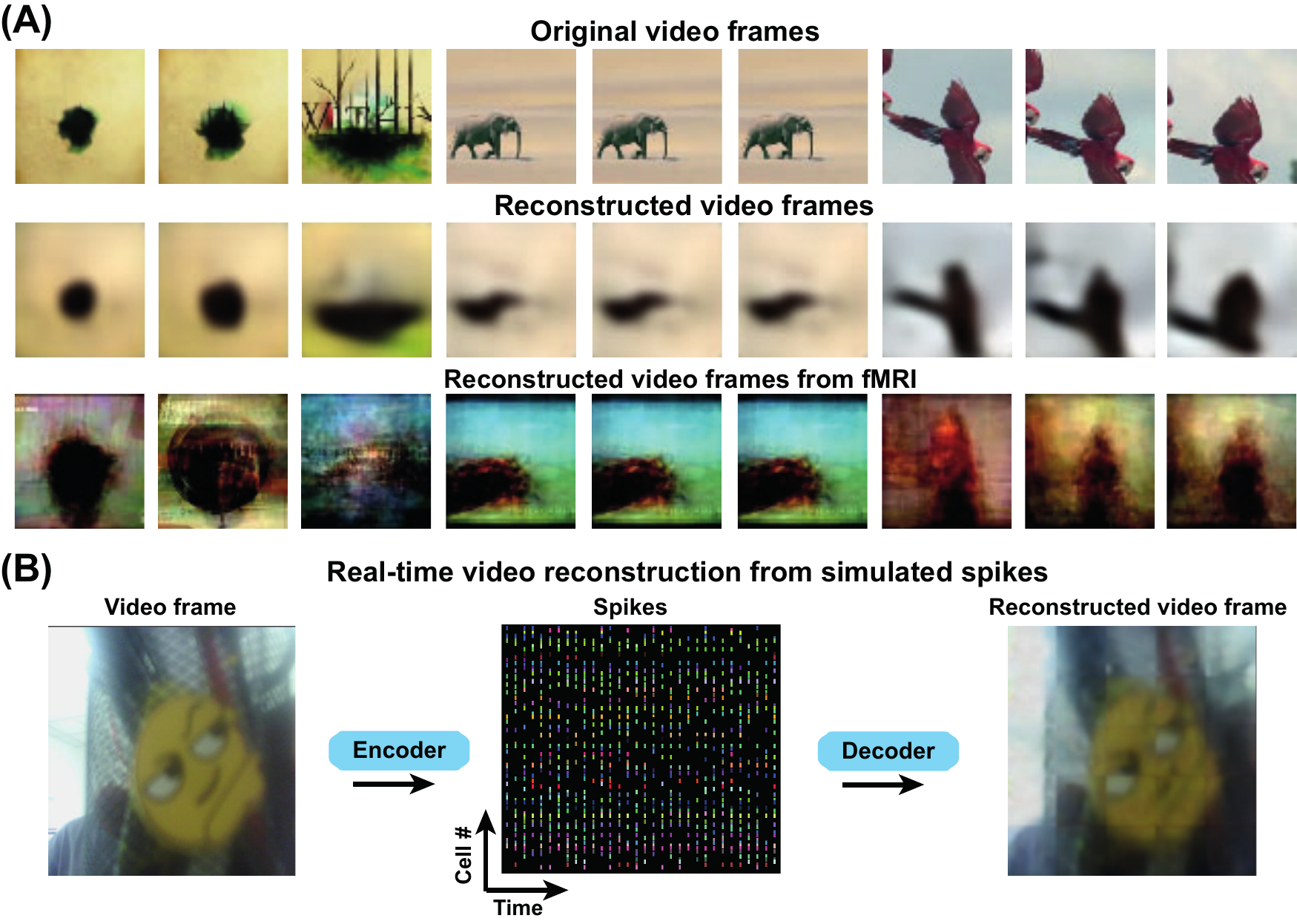}
	\end{center}
	\caption{ SID can be generalized to reconstruct arbitrary videos with a pre-trained model on CIFAR10 without any fine tunning. (A) Three segments of original videos (top),  reconstructed videos from simulated spikes with a SID model pre-trained on CIFAR10 data (middle), and  reconstructed videos based on fMRI signals from~\cite{nishimoto2011reconstructing} (bottom . (B) Decoding of real-time videos. (Left) One example frame of a real-time video. (Middle) A population of spike trains obtained by the encoder. Each line represents one RGC's spike train. (Right) The same frame reconstructed in a real-time fashion by the SID. Real-time videos are available to view online.} 
	\label{fig:model_movie}
\end{figure}

\subsection{Model generalization to real visual scenes}

Next, we used the encoding model to simulated RGCs in response to dynamic videos. For comparison, the same set of videos used in~\citep{nishimoto2011reconstructing} were encoded into RGC spikes. Then the SID model trained with CIFAR10 was used to reconstruct the videos from simulated RGC spikes. As shown in Fig~\ref{fig:model_movie} (A), the SID can reconstruct the dynamic videos from spikes very well.
Compared with the fMRI decoded results~\citep{nishimoto2011reconstructing}, our reconstructed videos contain more semantic contents and are much more refined. What is more, the SID model used here is trained with CIFAR10, which means that our SID model has a great generalization capability for arbitrary visual scenes. The complete videos reconstructed are available online (see Methods).

Finally, based on the results above, we implemented a real-time artificial vision system (see Methods), where real-time videos are captured by a standard camera, then encoded into a population of neural spikes, which are fed into the SID to reconstruct the visual scenes. Figure~\ref{fig:model_movie} (B) shows a snapshot of a real-time recorded video. Certainly, the speed of reconstruction and resolution of decoded video can be greatly improved with more advanced hardwares. For this real-time system, the SID is also trained with the same CIFAR-10 dataset. These results suggest that our model can sever as a general framework to analyze those data recorded by artificial event-based vision systems~\citep{guillermo2019event}.

\section{Discussion}

In this study, we proposed an end-to-end spike-image decoder to reconstruct stimulus images from neural spikes based on the retinal ganglion cells. The capability of SID was tested on experimental RGC data for both static and dynamic natural scenes to obtain state-of-the-art reconstruction performance. Furthermore, by using an additional encoding model, the great performance of reconstruction of arbitrary visual scenes was demonstrated on popular image datasets. In addition, our SID can also be used for decoding of fMRI data and show a very good performance compared to other methods developed in the context of fMRI decoding. Finally, we examined the generalization ability of SID by using a pre-trained SID to decode various types of dynamic videos to achieve real-time encoding and decoding visual scenes by neural spikes. 

\subsection{Rate code v.s. temporal code}
In our current study, the firing rate averaged over a number of trials for experimental data was used for our decoder and reconstruction. Both firing rate code and temporal code have been used in the field of neuronal coding~\cite{Jacobs2009, gollisch2008rapid, Onken_2016}. In practice, for analyzing experimental data, the difference between these two depends on many factors. In our case presented here, experimental data were collected under a condition that a 30 Hz frame rate was used for video stimulation, therefore, to match the sample rate of video frames, we binned the spike trains into 30 Hz as well. For image stimulation, each individual image was presented for 200 ms and then followed by an 800 ms empty display, therefore, for each image, we used the number of spikes during 300 ms after the onset of image presentation. We also conducted the reconstruction by binarizing the firing rate of RGCs into a single spike train with 0's and 1's only, and the results are similar to the current presentation.

From the experimental side, one could conduct new experiments with a high frame rate of video sampling, for instance, 120 Hz can be used, so that one only need to consider a smaller bin, 1000/120$\approx$8.3 ms, in which more precise spike times can be recovered given that the spare firing of the retinal ganglion cells in general can not generate 2 or more spikes within 8.3 ms in salamander, the animal species used here. On the other hand, these neurons in this animal species have a very low firing rate such that within a bin of 33 ms used in our study, there are 93.12\% of zero spike, 5.37\% of single spike, 1.14\% of double spikes, and 0.37\% of triple or more spikes. Therefore, although the firing rate was used for our decoding, the low firing rate in neurons is an approximation of temporal coding of spike times.

Given the experimental issues listed above, that is the reason why we use simulated spikes to demonstrate our method. In our simulations of image datasets and real-time videos, the decoding is based on the temporal code with the exact spikes in a single trial. Thus, our results presented here are provided as a proof-of-concept for using a population of single-trail spikes to reconstruct visual scenes. 

\subsection{Implication for other neuronal systems}

Our method could serve as a general model of neural coding, similar to other neural coding models, such as the general linear model~\cite{Pillow2008Spatio}, to study the relationship between neuronal signal and stimulus. In our case, the SID is a non-linear decoder to reconstruct image stimulus from neural spikes, which can be seen as an inverse function of spike-image mapping. Thus our method could be generalized to other parts of the brain as long as the neurons deliver different spike patterns for different stimuli. 

The retina has been thought of as a relatively simple neuronal circuit, however, it has been suggested that many complex computations can be realized by their neurons, in particular, the RGCs~\cite{Gollisch_2010}. It has been studied intensively to develop various types of models for the encoding of visual scenes in the retina~\cite{chichilnisky2001a, Pillow2008Spatio,Yu2020}. Similar to our current study, other recent work also focuses on neuronal spikes in the retina to develop the novel decoding methods~\cite{botella2018nonlinear,marre2015high, Parthasarathy2017}. All the methods used these studies take a similar approach by using the advantage of the recent advanced machine learning technique. Furthermore, the similar CNN decoding approach has been used for other parts of the visual system and different types of neural signals, such as calcium imaging in animal visual cortex~\cite{garasto2018visual}, and classical fMRI signal in human visual cortex~\cite{wen2018neural}. Therefore, our method could be applied to these types of neural signals from other parts of the visual system.

In the current study, we provide a case for using the SID to the fMRI signal. It is easy to expect that the decoding performance is higher for neural spikes compared to the fMRI signal due to the nature of neural signal, where spikes are much more refined than fMRI. The two-fold cross-validation shown in Figure~\ref{fig:fmri} confirm this expectation. With the same model, either SID or DGMM, the decoding results by using neural spikes are better than those by fMRI. However, interestingly, we also observe a methodological advantage by comparing two decoding methods: SID and DGMM. It should be noted that DGMM is a recent and state-of-the-art method developed in the context of fMRI decoding~\cite{Du2019}. Even in this condition, our SID can achieve a better decoding performance than DGMM by using the same fMRI data.
Therefore, our method can be seen as a more general model to cope with different types of neural signals beyond neural spikes.

Recent experimental advancements in neuroscience can collect a large population of neurons simultaneously. In particular, in the retina, a population of spike trains from hundreds of retinal ganglion cells can be obtained with well-controlled visual scenes, such as images and movies~\citep{botella2018nonlinear,Onken_2016}. New techniques may record several thousands of neurons simultaneously~\citep{portelli2016rank, maccione2014following,hilgen2017pan}. 
Together with other recent studies~\citep{botella2018nonlinear,Parthasarathy2017}, our results in this work show that a better decoding performance can be archived with neural spikes, which is beyond the scope of fMRI signal~\citep{wen2018neural, nishimoto2011reconstructing, Du2019}.

\subsection{Model limitation and further study}

Our decoder can be considered as a general framework, in which the detailed network structures are flexible. Here we only explored the simple and perhaps the most basic structure, e.g., a simple dense connected and layered network for the spike-image converter, similar to a multilayer perceptron,  and a typical network for the image-image autoencoder. One of the advantages of the current framework is that one can use any other types of spike-image converter, which is similar to existing models of decoding by transferring spikes into image pixels (see~\cite{botella2018nonlinear}), and any other types of image-image encoder, which is similar to existing models of image analysis (such as generative adversarial network or GAN~\citep{goodfellow2014generative}) in the deep learning field. Given the rapid development of architecture design for artificial, spiking, and a mixture of both~\cite{xu2018csnn}, neural networks, it is possible to use other network architectures to further improve the decoding ability. 

Our SID does not include the temporal filter, but only the spatial filter, i.e., the spatial receptive field of RGCs. 
RGCs have both spatial and temporal filters to take into account of spatial structure~\cite{Liu2017} and temporal dynamics of stimulus~\cite{Liu_2015}. Nonlinear structure in space has been intensively studied in the retina ~\cite{Liu2017, Schwartz2011, Schwartz2012}. In contrast, the temporal dynamics are more complex and entangled with the stimulus itself. Even for the simple artificial stimulus, such as white noise, there is a strong neuronal adaption although there is no correlation within white noise stimulus~\cite{Liu_2015}. For video stimulus with natural scenes, neuronal adaption entangled with correlation in the video in the temporal domain makes the investigation of neuronal coding of natural scene extremely difficult~\cite{alexander2016testing}. 

Here we overcome this difficulty by using a training set of randomly selected frames from the video in experimental data. In this way, one can break the temporal correlation between video frames. Essentially, this manipulation resembles presenting a large set of natural images sequentially, termed as naturalistic video, such that the temporal correlation between frames can be reduced to a minimal level~\cite{alexander2016testing}. The limited recording time during neuroscience experiments restricts the stimulus to a short segment of video used for RGCs. Thus, we tested the decoding capability of our method by using the stimulation of a large set of images, CIFAR, to mimic the full pixel space encountered by the model. To our surprise, our model trained by CIFAR can be generalized to any dynamic natural scenes without any re-training or fine-tunning. 

Dynamic visual scenes are highly complex with the information presented in a spatiotemporal fashion and high-order correlation~\citep{Simoncelli2001Natural}. Recent advancements of computer vision make some breakthroughs for analyzing these complex natural scenes~\citep{Lecun2015Deep}, for instance, the PredNet model can decode the next frame based on the previous frame of video~\cite{william2017deep}. 
However, the efficiency, generalization ability, and adaption or transfer learning between different tasks, of well-trained models are still far from human performance~\citep{Marblestone_2016}. Instead, our modeling framework shows a great generalization ability for the decoding of visual scenes with a pre-trained model. In future work, we expect to develop a time-based SID and use temporal filters for decoding.

\subsection{Implication for artificial visual systems}

The proposed model framework could also be used for visual neuroprostheses and other bioengineering devices of brain-machine interface. For the retinal neuroprostheses, the benefit of a decoding model is to justify the spiking patterns produced by the targeted downstream neurons, such that electrical stimulation should be able to close to those desired patterns of retinal neural activity in a prosthesis~\citep{Nirenberg2012}. Such control is important and beyond the traditional way of computing the distance between two or more spike trains in general~\citep{victor2005spike, rossum2001novel,shah2017learning}, as recent studies suggest that the performance of neuroprosthesis for restoring vision can be improved by the encoding/decoding algorithms besides of hardware design~\citep{Nirenberg2012, yue2016retinal, yan2018embedded}. Our study here proposes a novel decoding framework, as well as provides a proof-of-principle of encoding and decoding visual scenes with only a small population of neurons. Given the limitation of physical devices, such as only a small patch of brain area can be stimulated at one time, our framework suggests a feasible approach by using a small set of neurons for high precision decoding, which may systematically improve the capability of brain-machine interface.    

The model framework proposed in this study could be used for other artificial visual systems. The main feature of our framework is to make use of neural spikes. Advancements of recent artificial intelligence computing align with the development of the next generation of neuromorphic chips and devices, where the new data format is processed as spikes or events~\citep{guillermo2019event, Ma2017, Pei2019, esser2016convolutional,  davies2018loihi}. Therefore, our method can be applied to neuromorphic retina cameras with spiking or events, such as the dynamic vision sensor~\cite{lichtsteiner2008128}. Taken together with neuromorphic hardware and event/spiking computing algorithm, the next generation of computational vision can develop a better system for artificial vision.


\bibliographystyle{plain}
\bibliography{ref_all_new_nolink_liu}

\newpage

\text{\huge \centering Supplemental Materials:}

\text{\huge \centering
Reconstruction of Natural Visual Scenes }

\text{\huge \centering 
from Neural Spikes with Deep Neural Networks}

\renewcommand{\thefigure}{S\arabic{figure}}
\setcounter{figure}{0}

\renewcommand{\thetable}{S\arabic{table}}
\setcounter{table}{0}

\hspace{2cm}

\begin{figure*}[h]
	\begin{center}
		\includegraphics[width=0.95\columnwidth]{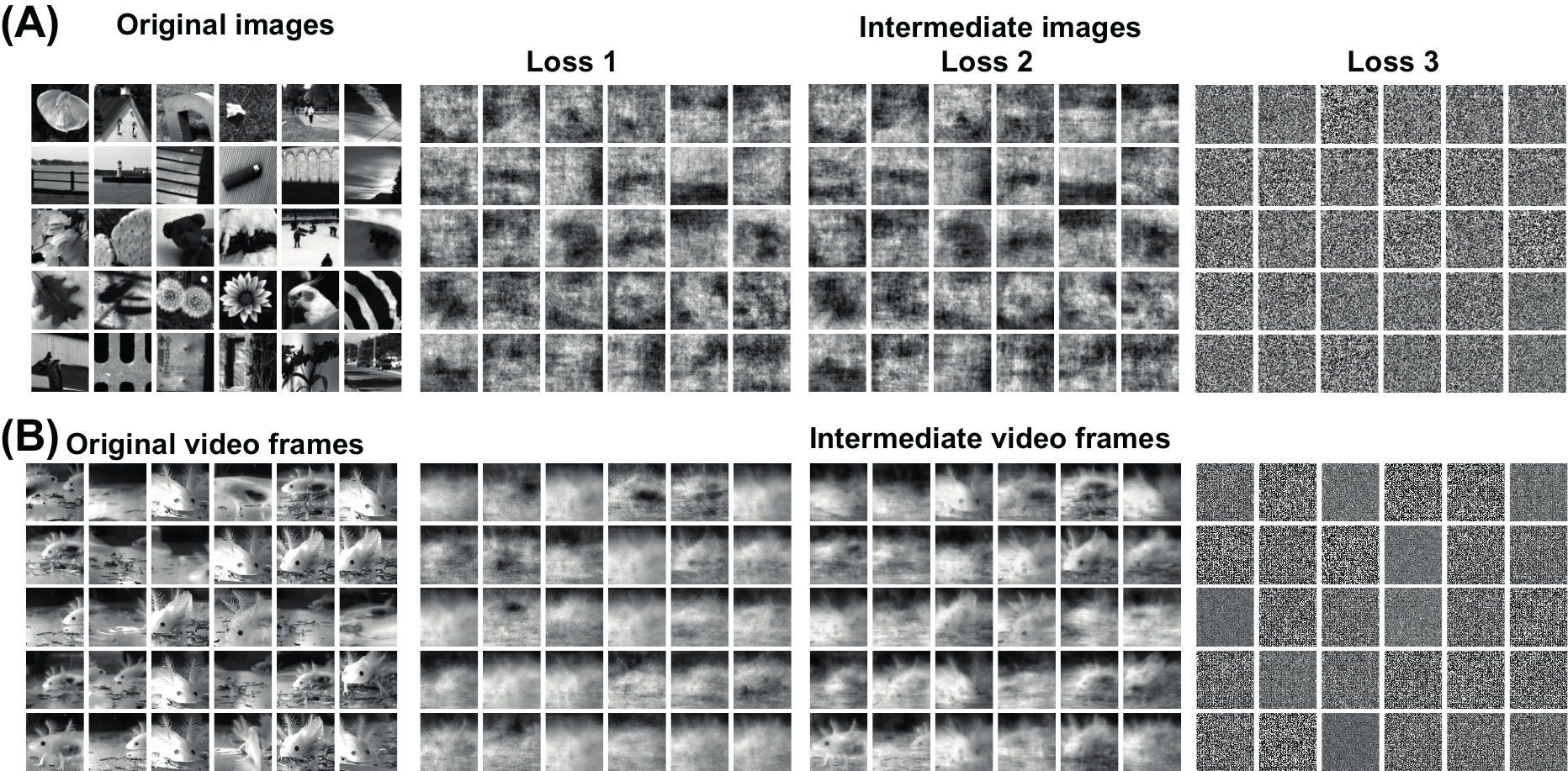}
	\end{center}
	\caption{Reconstruction from experimental RGC spikes with different loss functions. Original stimulus images in the test data. Intermediate images from RGC spikes with three loss functions. Different losses can lead to quite different intermediate images, in particular, random images for Loss 3.} 
	\label{fig:img_test_s}
\end{figure*}

\begin{figure*}[t]
	\begin{center}
		\includegraphics[width=0.95\columnwidth]{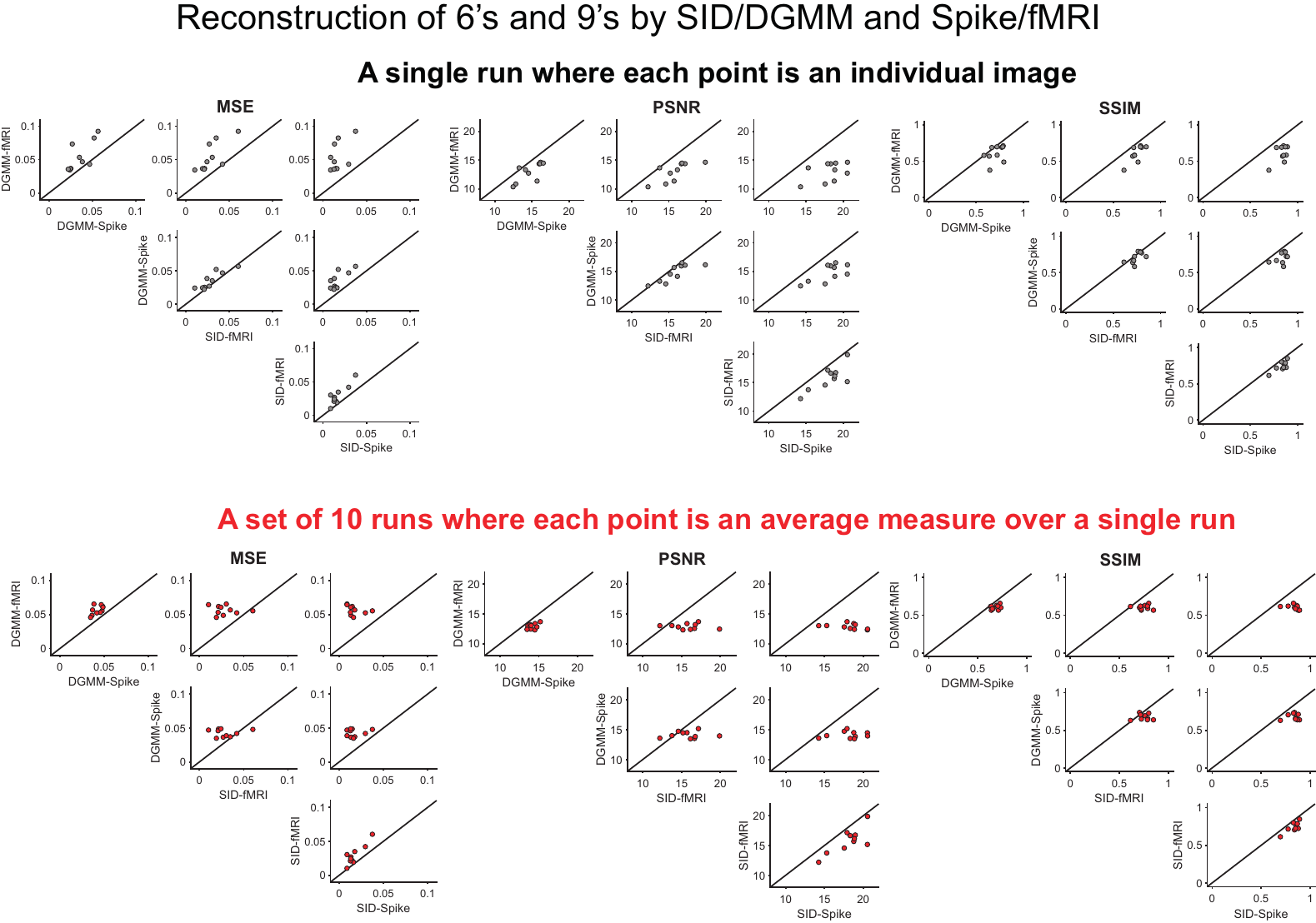}
	\end{center}
	\caption{Reconstruction from fMRI data and simulated spikes with the decoder of DGMM and SID cross-validated by a pair of model-data test, \textit{e.g.} DGMM-fMRI stands for fMRI decoding by the DGMM model. For comparison, cross-validation is done by using DGMM for simulated spikes and our SID for fMRI data. Model performance measured by MSE, PSNR, and SSIM in a single run of the SID model, and a set of 10 SID runs with different initial conditions. Each plot is a model-data pairwise comparison of MSE, PNSR, and SSIM. } 
	\label{fig:img_test_s}
\end{figure*}

\end{document}